\documentclass[reprint,aps,prc,amsmath,amssymb,superscriptaddress,preprintnumbers,floatfix]{revtex4-1}
\usepackage{graphicx}
\usepackage{dcolumn}
\usepackage{isotope}
\usepackage{hyperref}

\newcommand{\norm}[1]{\ensuremath{|\mathbf{#1}|}}

\begin{document}


\title{\texorpdfstring{Measurement of the cross sections for inclusive electron scattering\\ in the E12-14-012 experiment at Jefferson Lab}{Measurement of the cross sections for inclusive electron scattering in the E12-14-012 experiment at Jefferson Lab}}

\author{M.~Murphy} \affiliation{Center for Neutrino Physics, Virginia Tech, Blacksburg, Virginia 24061, USA}
\author{H.~Dai} \affiliation{Center for Neutrino Physics, Virginia Tech, Blacksburg, Virginia 24061, USA}
\author{L.~Gu} \affiliation{Center for Neutrino Physics, Virginia Tech, Blacksburg, Virginia 24061, USA}
\author{D.~Abrams} \affiliation{Department of Physics, University of Virginia, Charlottesville, Virginia 22904, USA}
\author{A.~M.~Ankowski}\email{ankowski@slac.stanford.edu} \affiliation{SLAC National Accelerator Laboratory, Stanford University, Menlo Park, California 94025, USA}
\author{B.~Aljawrneh} \affiliation{North Carolina Agricultural and Technical State University, Greensboro, North Carolina 27401, USA}
\author{S.~Alsalmi} \affiliation{Kent State University, Kent, Ohio 44242, USA}
\author{J.~Bane} \affiliation{The University of Tennessee, Knoxville, Tennessee 37996, USA}
\author{S.~Barcus} \affiliation{The College of William and Mary, Williamsburg, Virginia 23187, USA}
\author{O.~Benhar} \affiliation{INFN and Dipartimento di Fisica, Sapienza Universit\`{a} di Roma, I-00185 Roma, Italy}
\author{V.~Bellini} \affiliation{INFN, Sezione di Catania, Catania, 95123, Italy}
\author{J.~Bericic} \affiliation{Thomas Jefferson National Accelerator Facility, Newport News, Virginia 23606, USA}
\author{D.~Biswas} \affiliation{Hampton University, Hampton, Virginia 23669, USA}
\author{A.~Camsonne} \affiliation{Thomas Jefferson National Accelerator Facility, Newport News, Virginia 23606, USA}
\author{J.~Castellanos} \affiliation{Florida International University, Miami, Florida 33181, USA}
\author{J.-P.~Chen} \affiliation{Thomas Jefferson National Accelerator Facility, Newport News, Virginia 23606, USA}
\author{M.~E.~Christy} \affiliation{Hampton University, Hampton, Virginia 23669, USA}
\author{K.~Craycraft} \affiliation{The University of Tennessee, Knoxville, Tennessee 37996, USA}
\author{R.~Cruz-Torres} \affiliation{Massachusetts Institute of Technology, Cambridge, Massachusetts 02139, USA}
\author{D.~Day} \affiliation{Department of Physics, University of Virginia, Charlottesville, Virginia 22904, USA}
\author{S.-C.~Dusa} \affiliation{Thomas Jefferson National Accelerator Facility, Newport News, Virginia 23606, USA}
\author{E.~Fuchey} \affiliation{University of Connecticut, Storrs, Connecticut 06269, USA}
\author{T.~Gautam} \affiliation{Hampton University, Hampton, Virginia 23669, USA}
\author{C.~Giusti} \affiliation{Dipartimento di Fisica, Universit\`{a} degli Studi di Pavia and INFN, Sezione di Pavia,  I-27100 Pavia, Italy}
\author{J.~Gomez} \affiliation{Thomas Jefferson National Accelerator Facility, Newport News, Virginia 23606, USA}
\author{C.~Gu} \affiliation{Duke University, Durham, North Carolina 27708, USA}
\author{T.~Hague} \affiliation{Kent State University, Kent, Ohio 44242, USA}
\author{J.-O.~Hansen} \affiliation{Thomas Jefferson National Accelerator Facility, Newport News, Virginia 23606, USA}
\author{F.~Hauenstein} \affiliation{Old Dominion University, Norfolk, Virginia 23529, USA}
\author{D.~W.~Higinbotham} \affiliation{Thomas Jefferson National Accelerator Facility, Newport News, Virginia 23606, USA}
\author{C.~Hyde} \affiliation{Old Dominion University, Norfolk, Virginia 23529, USA}
\author{C.~M.~Jen} \affiliation{Center for Neutrino Physics, Virginia Tech, Blacksburg, Virginia 24061, USA}
\author{C.~Keppel} \affiliation{Thomas Jefferson National Accelerator Facility, Newport News, Virginia 23606, USA}
\author{S.~Li} \affiliation{University of New Hampshire, Durham, New Hampshire 03824, USA}
\author{R.~Lindgren} \affiliation{Department of Physics, University of Virginia, Charlottesville, Virginia 22904, USA}
\author{H.~Liu} \affiliation{Columbia University, New York, New York 10027, USA}
\author{C.~Mariani} \affiliation{Center for Neutrino Physics, Virginia Tech, Blacksburg, Virginia 24061, USA}
\author{R.~E.~McClellan} \affiliation{Thomas Jefferson National Accelerator Facility, Newport News, Virginia 23606, USA}
\author{D.~Meekins} \affiliation{Thomas Jefferson National Accelerator Facility, Newport News, Virginia 23606, USA}
\author{R.~Michaels} \affiliation{Thomas Jefferson National Accelerator Facility, Newport News, Virginia 23606, USA}
\author{M.~Mihovilovic} \affiliation{Jozef Stefan Institute, Ljubljana 1000, Slovenia}
\author{D.~Nguyen} \affiliation{Department of Physics, University of Virginia, Charlottesville, Virginia 22904, USA}
\author{M.~Nycz} \affiliation{Kent State University, Kent, Ohio 44242, USA}
\author{L.~Ou} \affiliation{Massachusetts Institute of Technology, Cambridge, Massachusetts 02139, USA}
\author{B.~Pandey} \affiliation{Hampton University, Hampton, Virginia 23669, USA}
\author{V.~Pandey} \affiliation{Center for Neutrino Physics, Virginia Tech, Blacksburg, Virginia 24061, USA}
\author{K.~Park} \affiliation{Thomas Jefferson National Accelerator Facility, Newport News, Virginia 23606, USA}
\author{G.~Perera} \affiliation{Department of Physics, University of Virginia, Charlottesville, Virginia 22904, USA}
\author{A.~J.~R.~Puckett} \affiliation{University of Connecticut, Storrs, Connecticut 06269, USA}
\author{S.~N.~Santiesteban} \affiliation{University of New Hampshire, Durham, New Hampshire 03824, USA}
\author{S.~\v{S}irca} \affiliation{University of Ljubljana, Ljubljana 1000, Slovenia} \affiliation{Jozef Stefan Institute, Ljubljana 1000, Slovenia}
\author{T.~Su} \affiliation{Kent State University, Kent, Ohio 44242, USA}
\author{L.~Tang} \affiliation{Hampton University, Hampton, Virginia 23669, USA}
\author{Y.~Tian} \affiliation{Shandong University, Shandong, 250000, China}
\author{N.~Ton} \affiliation{Department of Physics, University of Virginia, Charlottesville, Virginia 22904, USA}
\author{B.~Wojtsekhowski} \affiliation{Thomas Jefferson National Accelerator Facility, Newport News, Virginia 23606, USA}
\author{S.~Wood} \affiliation{Thomas Jefferson National Accelerator Facility, Newport News, Virginia 23606, USA}
\author{Z.~Ye} \affiliation{Physics Division, Argonne National Laboratory, Argonne, Illinois 60439, USA}
\author{J.~Zhang} \affiliation{Department of Physics, University of Virginia, Charlottesville, Virginia 22904, USA}

\collaboration{The Jefferson Lab Hall A Collaboration}\noaffiliation


\begin{abstract}
The E12-14-012 experiment performed at Jefferson Lab Hall A has collected inclusive electron-scattering data for different targets at the kinematics corresponding to beam energy 2.222~GeV and scattering angle $15.54^\circ$. Here we present a comprehensive analysis of the collected data and compare the double-differential cross sections for inclusive scattering of electrons, extracted using solid targets (aluminum, carbon, and titanium) and a closed argon-gas cell. The data extend over broad range of energy transfer, where quasielastic interaction, $\Delta$-resonance excitation, and inelastic scattering yield contributions to the cross section. The double-differential cross sections are reported with high precision ($\sim$3\%) for all targets over the covered kinematic range.
\end{abstract}

\preprint{JLAB-PHY-19-3013}
\preprint{SLAC-PUB-17457}

\maketitle
\section{Introduction}
Electron-scattering experiments have been shown to be the best tool for precise investigations of the structure of atomic nuclei~\cite{Boffi:1996ikg}. The electromagnetic interaction of electrons with the target is weak, compared with the strength of interactions that bind nucleons together, and can be treated as an exchange of a single photon. Allowing the nuclear response to be probed at energy transfers varied independently from momentum transfers, electron beams can be used to investigate physics corresponding to various excitation energies with different spatial resolutions, exposing to different interaction mechanisms.

The existing body of electron-scattering data clearly shows that many important features of nuclear structure can be described by assuming that nucleons forming the nucleus behave as independent particles bound in a mean field~\cite{deWittHuberts:1990zy}, but this picture is not complete without accounting for correlations between nucleons~\cite{Rohe:2004dz,degliatti:2015,Fomin:2017ydn}.

While analysis of electron scattering from nuclei is interesting in its own right, accurate description of nuclear effects in interactions of few-GeV probes is now coming into sharp focus due to its relevance for neutrino physics. As neutrino oscillation parameters are extracted from collected event spectra, and neutrino energies have to be reconstructed from the observed kinematics of the products of their interactions with nuclear targets, nuclear effects play a fundamental role in neutrino-oscillation analysis~\cite{Ankowski:2016jdd}.

In neutrino experiments, nuclear models implemented in Monte Carlo (MC) simulations are employed to predict event rate in a near detector, perform its extrapolation to a far detector, estimate the energy carried by undetected particles, and obtain background estimates. While description of nuclear effects is already one of the largest sources of systematic uncertainties in ongoing oscillation studies~\cite{Abe:2018wpn}, its importance will increase further in the next generation of oscillation experiments. In particular, to achieve their sensitivity goals, the Deep Underground Neutrino Experiment (DUNE) and Hyper-Kamiokande have to reduce uncertainties coming from nuclear cross sections to the few-percent level~\cite{Acciarri:2015uup,Abe:2015zbg}.

As weak interactions of neutrinos probe the nucleus in a very similar way as electromagnetic interactions of electrons, precise electron-scattering data give unique opportunity to validate nuclear models employed in neutrino physics. A theoretical model unable to reproduce electron measurements cannot be expected to provide accurate predictions for neutrino cross sections.

At the kinematics where the impulse approximation is valid---the process of scattering off a nuclear target can be described as involving predominantly a single nucleon, with $(A - 1)$ nucleons acting as a spectator system---nuclear effects can be separated from the description of the elementary cross sections, differing between neutrinos and electrons, and the knowledge gained in electron scattering directly translates to neutrino interactions. In particular, measurements of the $(e,e'p)$ cross sections---in which knocked out protons are detected in coincidence with electrons---can be used to extract the information on the momentum and energy distributions (the spectral function) of protons in the nucleus, and on final-state interactions (FSI) of the struck protons propagating through the (excited) residual nucleus, which are intrinsic properties of the target and do not depend on the interaction mechanism.

In the simplest case of a symmetric nuclear target, with the proton number $Z$ equal to the neutron number $N$, nuclear effects are expected to be largely the same in neutrino and electron interactions, up to small Coulomb corrections. For an asymmetric nucleus, one needs to additionally analyze electron scattering on its mirror nucleus, with $Z$ and $N$ swapped, to obtain a good approximation of information on the neutron structure, impossible to collect directly. In the case of DUNE, in addition to argon ($Z=18, N=22$)---employed as the target material---it is necessary to collect electron-scattering data also for titanium ($Z=22, N\simeq26$). While the exclusive $(e,e'p)$ cross sections give direct insight into the nuclear structure, they do not provide a complete picture of all interaction dynamics.

When the energy transferred by the interacting electron to the nucleon increases, the interaction mechanism changes from quasielastic (QE) scattering, in which the struck nucleon is removed from the nucleus, to nucleon resonance production---dominated by the excitation of the $\Delta$ resonance---and finally to deep-inelastic scattering (DIS) on individual quarks forming the nucleon.

The inclusive $(e,e')$ measurements, which yield the spectra of electrons scattered at fixed angle, provide information on all interaction mechanisms, regardless of the composition of hadrons in the final state. As a consequence, a great deal can be learned from the inclusive $(e,e')$ cross sections, particularly in the context of DUNE, in which $\sim$2/3 of events are expected to involve pions~\cite{Acciarri:2015uup}.

The features of the peaks observed in the in{\-}clu{\-}sive spectrum---their width, position, shape, and height---provide information on the momentum and energy distributions of the nucleons in the nuclear ground state, as well as on the final-state interactions between the struck-nucleons and the spectator system.

The width of the QE peak, which in the nonrelativistic regime depends on both the momentum carried by the struck nucleon and the momentum transfer, ${\bf q}$, in the relativistic regime becomes largely independent of ${\bf q}$, and can be simply parametrized in terms of a Fermi momentum, $k_F$~\cite{Moniz:1971mt}. However, a kinematics-dependent broadening ascribed to FSI is also observed.

The position of the QE peak is determined by the combined effects of nuclear binding and FSI,
and the height of the QE peak depends on the number of nucleons probed by the interaction, and the momentum and energy distributions of nucleons in the ground state.

The identification of nuclear effects shaping the peak corresponding to QE scattering largely applies to other interaction mechanisms as well. However, their contributions give rise to broader structures in the cross section, as they involve production of hadrons of finite lifetimes.

To provide a reliable foundation for the oscillation analysis of precise long-baseline neutrino experiment, a~nuclear model employed in Monte Carlo simulations must be able to reproduce the features of the cross sections for electron scattering corresponding to the kinematics and target of relevance. In the context of DUNE, beam energies between 2 and 4~GeV play the most important role, and argon is the target material.

Previously~\cite{Dai:2018gch,Dai:2018xhi}, we reported the inclusive cross sections for electron scattering off argon, titanium, and carbon, extracted for beam energy 2.222~GeV and scattering angle $15.54^\circ$. Here we present a new result for aluminum, and a complete scaling analysis of all the targets that we have analyzed. We also discuss in more details the procedures used to measure the inclusive cross sections, together with their uncertainty estimates. In Sec.~\ref{sec:ExperimentalSetup} we describe the setup of the performed experiment. The methods of extracting the cross sections are presented in Sec.~\ref{sec:DataAnalysis}. The estimates of their uncertainties are covered in Sec.~\ref{sec:Uncertainties}. In Sec.~\ref{sec:Results} we report and discuss the measured aluminum cross section, while Sec.~\ref{sec:scaling} is devoted to the scaling analysis of our data. Finally, in Sec.~\ref{sec:Summary} we summarize our findings and draw the conclusions.
%
%
%
%
%
%
%
%
\section{Experimental Setup}\label{sec:ExperimentalSetup}
Performed at Jefferson Lab, the experiment E12-14-012 took both exclusive  $(e,e'p)$ electron-scattering data---in which the proton knocked out from the nuclear target is detected in coincidence with the scattered electron---and inclusive $(e,e')$ data---in which all final states contribute---for different targets: C, Ti, Ar, and Al. The data analysis for inclusive electron scattering was relatively simple, as it implied modest data-acquisition (DAQ) rates and very small pion backgrounds. E12-14-012 used an electron beam of energy 2.222~GeV provided by the Continuous Electron Beam Accelerator Facility (CEBAF), and took data in Spring 2017. The average beam current was 10~$\mu$A. Scattered electrons were measured by using a high resolution spectrometer (HRS), equipped with two vertical drift chambers (VDCs) providing tracking information~\cite{Fissum:2001st}, two scintillator planes for timing measurements and triggering, double-layered lead-glass calorimeter, and a gas \v{C}erenkov counter used for particle identification~\cite{Alcorn:2004sb}. The HRS was positioned with a central scattering angle of $\theta=15.54^\circ$. The beam current and position, the latter being critical for the electron-vertex reconstruction and momentum calculation, were monitored by resonant radio-frequency cavities (beam current monitors, or BCMs~\cite{Alcorn:2004sb}) and cavities with four antennae (beam position monitors, or BPMs~\cite{Alcorn:2004sb}), respectively. The beam size was measured by using harp scanners, which moved a thin wire through the beam. The beam was spread over a $2\times2$ mm$^2$ area to avoid overheating the target.

The experiment employed a set of solid targets---aluminum, carbon (single foil and a multifoil composed of nine foils), and titanium---as well as a closed cell of gaseous argon~\cite{Santiesteban:2018qwi}. The aluminum target consisted of two identical foils of the 7075 alloy, the thickness of which was $0.889\pm0.002$~g/cm$^2$. Details of the elementary composition of the Al-7075 alloy used in the E12-14-012 experiment are given in Table~\ref{tab:alcomp}. The aluminum foils were positioned to match the entrance and exit windows of the argon target, separated by a distance of 25~cm. Their thickness was fixed to be the same as the radiation length of the argon target. The analysis presented here uses the data from one of the foils only, located upstream of the spectrometers at $z=-12.5$~cm. The data were taken in nine separate runs, modifying at each step the momentum of the spectrometer in order to cover the final electron energy $E^\prime$ from 1.285 to 2.135~GeV.

\par The VDCs' tracking information allowed the determination of the momentum and reconstruction of the direction (in-plane and out-of-plane angles) of the scattered electron, and reconstructing the interaction vertex at the target. The transformation between focal plane and target quantities was computed by using an optical matrix, the accuracy of which was verified by using the multifoil target data and sieve measurements. Possible variations of the magnetic field in the HRS magnets that could affect the optics are included in the analysis as systematic uncertainties.

We set up two types of hardware triggers:
\[\begin{split}
T_3 &= (S_0 \&\& S_2) \&\& (LC || GC),\\
T_5 &= (S_0 || S_2) \&\& (LC || GC).\\
\end{split}\]
The $T_3$ ($T_5$) trigger type requires that the signal from the scintillator plane $S_0$ {\sc and} $S_2$ ($S_0$ {\sc or} $S_2$) is detected in coincidence with the signal from the lead calorimeter ($LC$) {\sc or} the gas \v{C}erenkov counter ($GC$).

\begin{table}
\caption{\label{tab:alcomp}Composition of the Al-7075 alloy. For each element, we provide the number of protons, $Z$, and the average number of neutrons, $N$, calculated according to the isotopic abundances~\cite{Tuli:2011}.}
\begin{ruledtabular}
\begin{tabular}{@{}c d d d@{}}
   & \multicolumn{1}{c}{weight (\%)} & Z & N \\
\hline
Al & 89.72 & 13 & 14.00 \\
Zn &  5.8  & 30 & 35.45 \\
Mg &  2.4  & 12 & 12.32 \\
Cu &  1.5  & 29 & 34.62 \\
Fe &  0.19 & 26 & 29.91 \\
Cr &  0.19 & 24 & 28.06 \\
Si &  0.07 & 14 & 14.11 \\
Mn &  0.03 & 25 & 30.00 \\
Ti &  0.03 & 22 & 25.92 \\
V  &  0.01 & 23 & 28.00 \\
Zr &  0.01 & 40 & 51.32 \\
other &  0.05 & & \\
\hline
average & & \multicolumn{1}{c}{$14.26\pm0.01$} & \multicolumn{1}{c}{$15.58\pm0.01$}\\
\end{tabular}
\end{ruledtabular}
\end{table}
Electrons were selected in the HRS requiring, in addition, one reconstructed track. Furthermore, they had to deposit 30\% of their energy in the lead calorimeter (${E_\text{cal}}/{p} > 0.3$) and had a signal in the \v{C}erenkov counter of more than 400 analog-digital-converter (ADC) counts.  To select events in the central acceptance region of the spectrometer, the electron's track was required to be within $\pm$4~mrad of the in-plane angle and $\pm$6~mrad of the out-of-plane angle with respect to the center ray of the spectrometer and have a $dp/p$ of $\pm$0.04. The spectrometers were calibrated by using sieve slit measurements and the position of the spectrometers and angles were surveyed before the start of the data taking. The survey precision for the angle and position measurements is respectively 0.01 mrad and 0.01 mm.
\par The efficiencies of the elements in the detector stack were studied by comparing rates in various combinations of secondary triggers. The scintillator efficiency, $(S_0\&\&S_2)$, was studied by using the ratio of the events rates selected using $T_3$ and $T_5$ trigger types, requiring one reconstructed track, and applying the acceptance and calorimeter cuts. It was found to be 99\%. The efficiency of the calorimeters was close to 100\% for all runs, and the efficiency of the \v{C}erenkov detector was found to range from 99.9\% for the highest $E^\prime$ runs down to 97.5\% for the lowest $E^\prime$ run. The \v{C}erenkov efficiency was evaluated by selecting a pure sample of electrons in the calorimeter and varying the \v{C}erenkov cut by $\pm$10\%. A summary of the efficiency is presented in Table~\ref{tab:eff}. There is a large variation of the tracking efficiency between the QE and DIS regimes due to the requirement of only one reconstructed track.

The livetime of the electronics was computed by using the rates from scalers, which were independent of triggered events. The acceptance-cut efficiencies and the $dp$ cut efficiency were computed by using the MC simulation~\cite{Arrington:1999}. As the efficiencies in Table~\ref{tab:eff} turn out to be very similar for the different kinematic regions of the inclusive cross section, we report their ranges of variation.
The overall efficiency (between 83\% and 95\% across all the kinematic regions) includes cuts on the calorimeters, both the lead and the \v{C}erenkov counter, track reconstruction efficiency, livetime, and acceptance.

\begin{table}
\caption{\label{tab:eff}Efficiencies used to calculate Al yield, reported as variation ranges across all the kinematic regions.}
\begin{ruledtabular}
\begin{tabular}{@{}ccr l lccc@{}}
	&&	&							&& Efficiency 	&&  \\
\hline				
&&a. & Livetime && 	  97.8--98.6\% &&\\
&&b. & Tracking  &&	  88.0--94.9\% &&\\
&&c. & Trigger &&	  99.2--99.6\% &&\\
&&d. &\v{C}erenkov && 97.5--99.9\% &&\\
&&e. &Calorimeter &&  99.8--99.9\% &&\\
\end{tabular}
\end{ruledtabular}
\end{table}
\section{Data Analysis}\label{sec:DataAnalysis}
\subsection{Yield-Ratio Method}\label{sec:yield}
The yield-ratio method of determining the cross section involves both the experimental data and simulated MC data. In this method, the yield $Y$ is calculated for both sets of data as
\begin{equation}
Y^i=(N_S^i \times PS) / (LT \times \epsilon),
\label{eq:yield}
\end{equation}
where $i$ refers to the $i${th} bin of the $E^\prime$ distribution, $N^i_S$ is the total number of scattered electrons, $PS$ is a prescale factor in the DAQ, $\epsilon$ is the total efficiency of the detector, and $LT$ is the livetime of the electronics. The ratio of the yields for the actual and MC data is taken as a means of eliminating any impact of the acceptance on each bin, and then the measured cross section is determined by multiplying this ratio by the cross section calculated within the Monte Carlo model:
\begin{equation}
\frac{d^{2}\sigma_\text{data}}{d\Omega dE'}=\frac{d^{2}\sigma_\text{MC}}{d\Omega dE'}  \frac{Y_\text{data}}{Y_\text{MC}}.
\end{equation}
The MC cross section is a fit to existing data, including preliminary Hall C data, accounting for radiative corrections computed by using the peaking approximation~\cite{Whitlow:1990} and Coulomb corrections implemented with an effective momentum approximation~\cite{Aste:2005wc}.
\subsection{Acceptance Method}\label{sec:acceptance}
The cross section was also extracted via another method, the acceptance method, and both the yield ratio and acceptance methods were examined for agreement. In the case of the argon and titanium targets, for which accurate models of the nuclear response are not yet available, it is important to validate the MC simulation and results obtained using the yield-ratio method using an alternative method, less dependent on the input MC cross-section model. In this analysis, the acceptance method is meant to serve as a cross-check for potential biases of the yield-ratio method stemming from the employed cross-section model, in addition to the direct determination of its sensitivity by variations of the cross section. The acceptance method will be described in the following of this section.

For each  $(\Delta E, \Delta \Omega)$ bin, the number of detected electrons can be determined by using
\begin{equation}
N_S^i = L \frac{d^2\sigma}{d\Omega dE^\prime} \Delta E^\prime \Delta \Omega\,\epsilon A^i(E^\prime,\theta, \phi)
\end{equation}
where $L$ is the integrated luminosity (number of beam electrons $\times$ number of targets / area), $\epsilon$ is the total detection efficiency, and $\theta$ and $\phi$ represent the in-plane and out-of-plane angles, respectively. The acceptance $A^i(E^\prime,\theta, \phi)$ is the probability that a particle passes through the spectrometer into the $i$th bin, generated separately for each momentum setting of the spectrometer.

The electron yield corrected for the overall efficiency (product of individual efficiencies as described above) can be cast as
\begin{equation}\begin{split}
Y^i = \frac{N_S^i}{\epsilon}= L \frac{d^{2}\sigma_\text{data}}{d\Omega dE^\prime} \Delta E^\prime \Delta \Omega A^i(E^\prime,\theta, \phi),
\end{split}\end{equation}
and the cross section can be measured using
\begin{equation}
\frac{d^{2}\sigma_\text{data}}{d\Omega dE^\prime}=\frac{Y^i}{\Delta E^\prime \Delta \Omega A^i(E^\prime,\theta, \phi) L}.
\end{equation}

The single-arm Monte Carlo simulation was used to generate events uniformly distributed in $(\theta, \phi, E^\prime)$. For a specific phase-space slice in $(\Delta \theta, \Delta \phi, \Delta E^\prime)$, we computed the ratio between the total number of events that reach the spectrometer and the number of generated events. The ratio of these two numbers represents the probability that a particle successfully passes through the magnets and the aperture to arrive at the detector package.

For an extended target, an acceptance matrix $A^i(E^\prime,\theta, \phi)$ was generated at various points along the target length. Each different target slice was associated with a different $A^i(E^\prime,\theta, \phi)$.

The number and size of the slices were optimized based on the statistics of the data. In principle, an infinite number of matrices could be used in order to make events perfectly weighted, but this method would be inefficient and subject to large statistical fluctuations, if the number of events in each region was limited.

In this analysis, we used a single matrix for events along the entire target length to correct the data, and evaluated the residual variation along the beam direction~$z$. For these studies we took advantage of the optical target data, collected in the spring of 2017.

\begin{figure}
\centering
\includegraphics[width=0.9\columnwidth]{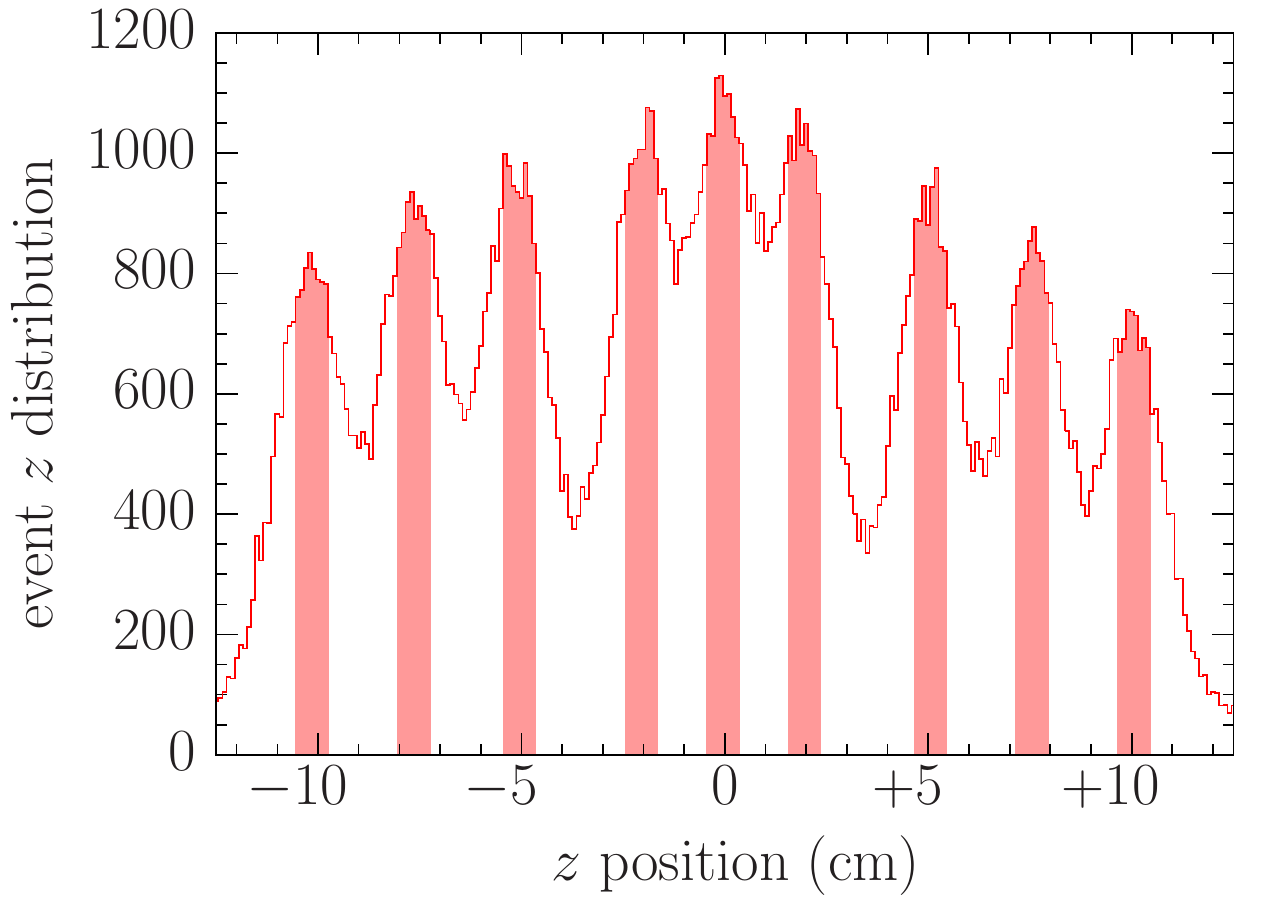}
\caption{(color online). Distribution along the beam direction of reconstructed events for the multifoil carbon target. The shaded regions represent the data selected to identify the events coming from individual carbon foils.}
\label{fig:optical_z_distribution}
\end{figure}

The optical target was a series of nine carbon foils, placed along the beam direction at $z=0$~cm, $\pm$2~cm, $\pm$5~cm, $\pm$7.5~cm, $\pm$10~cm. The $z$ distribution of the events reconstructed from the optical target is shown in Fig.~\ref{fig:optical_z_distribution}, with the shaded regions representing the $z$-position cuts employed to identify the events coming from individual carbon foils. Because it would be difficult to select pure events from each foil, due to its finite thickness, we used the Monte Carlo simulation and the carbon cross-section model to generate single-foil carbon data for different $z$ positions of the target.

\begin{figure}
\centering
\includegraphics[width=0.9\columnwidth]{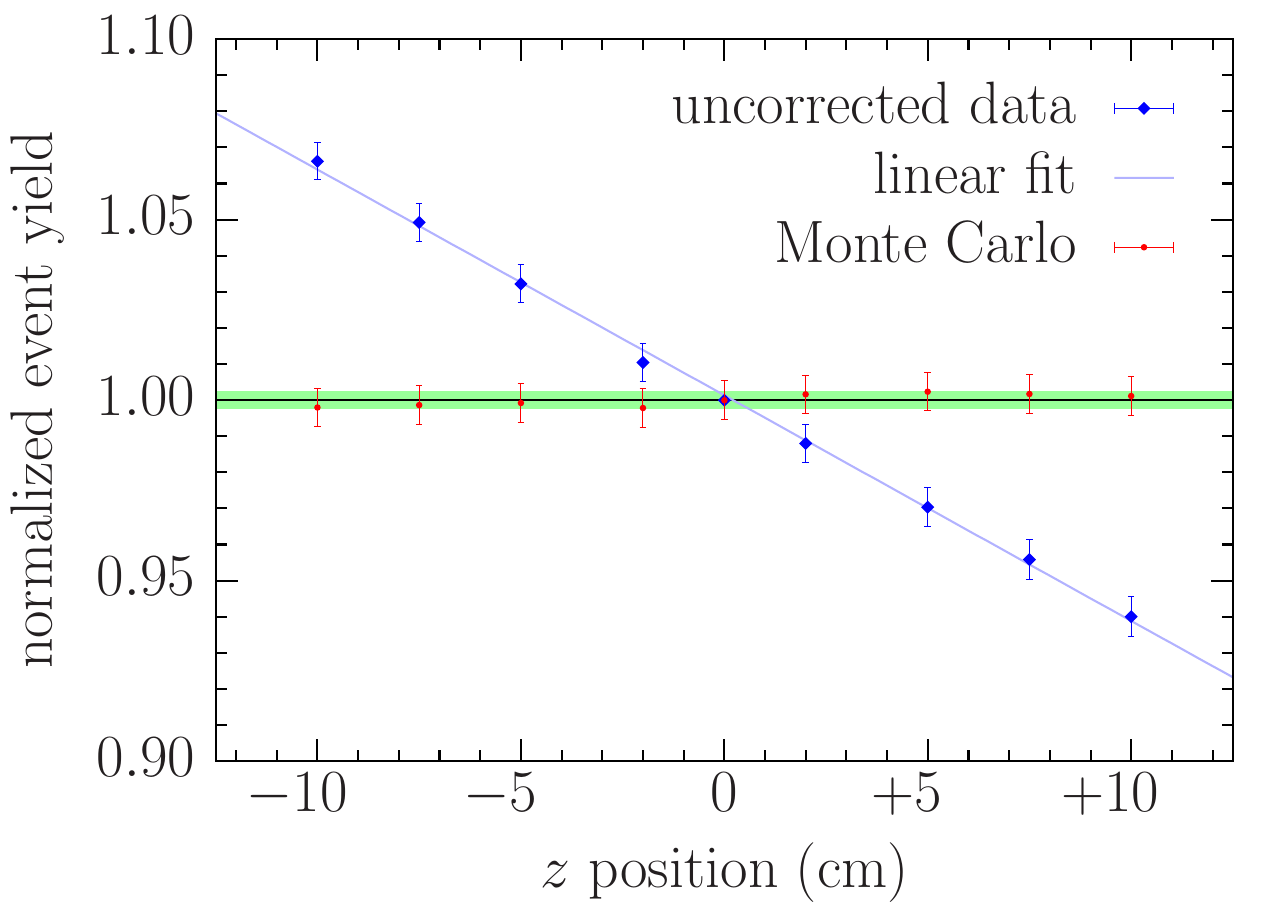}
\caption{(color online). Event yields from carbon foils at different positions along the beam direction, normalized to the yield for the central foil, for the uncorrected data and for the Monte Carlo simulation. The dependence of the cross section on the scattering angle, correctly taken into account in the Monte Carlo simulation, introduces a linear trend in the data that needs to be corrected for. All uncertainties are purely statistical.}
\label{fig:ratio_optical_foil_MC}
\end{figure}

Using the single-foil carbon data, we generated for each of the nine momentum settings a set of acceptance matrices corresponding to the mean $z$ position of each foil composing  the multifoil carbon target. We applied a weight of $1/A(E^\prime, \theta,\phi)$ to every event, and made a comparison between the events originating from individual foils. The obtained distribution of MC event yields from different foils, normalized to the one from the foil at $z=0$~cm, is shown in Fig.~\ref{fig:ratio_optical_foil_MC}. The results for the nine regions, represented by the red dots lying inside the green shaded band, are in excellent agreement, with variations between them remaining below 0.5\%.

\begin{figure}
\centering
\includegraphics[width=0.9\columnwidth]{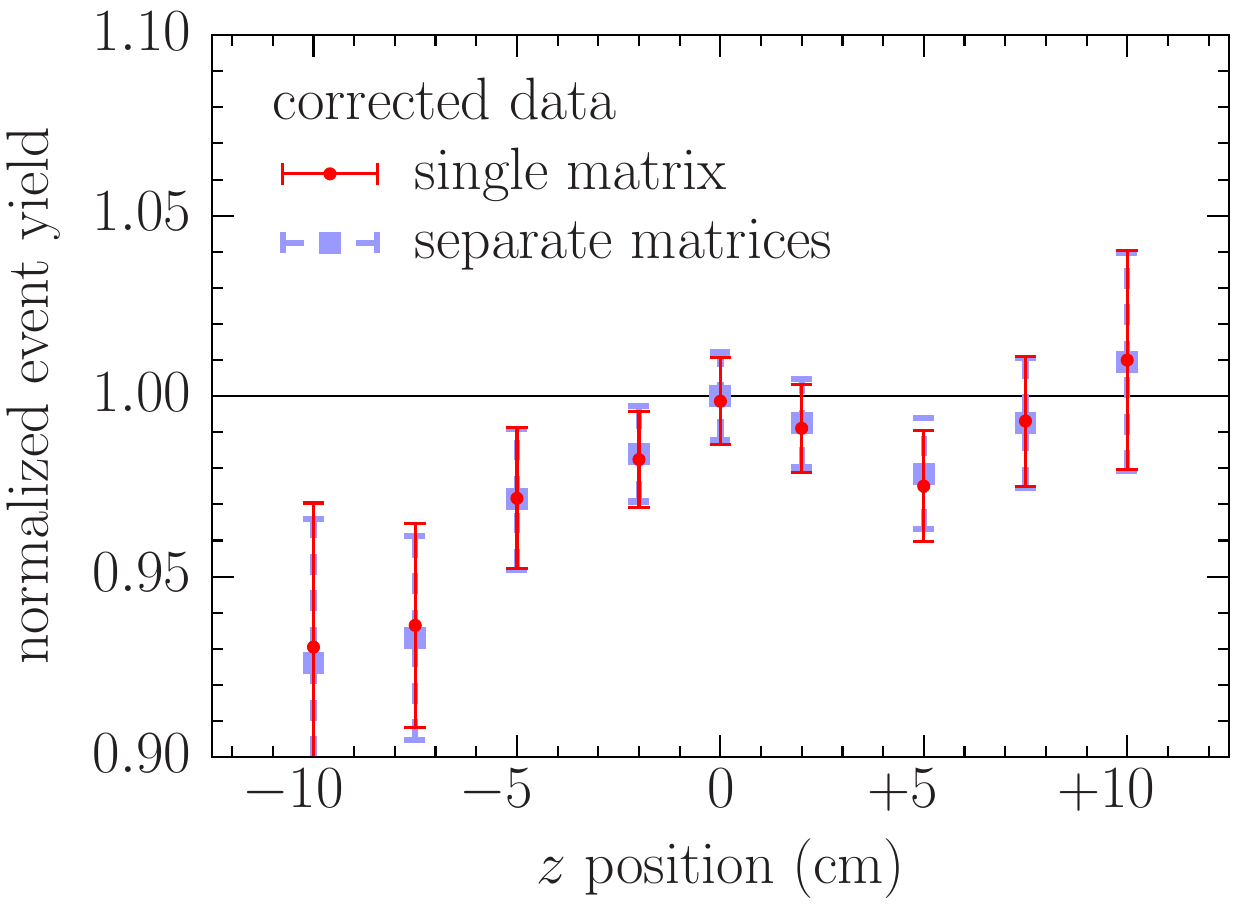}
\caption{(color online). Event yields in the corrected data for the multi-foil carbon target surviving the $z$-position selection, normalized to the yield for the central foil. The outcomes of two correction procedures are compared. The error bars are symmetric and represent the total uncertainties, being the statistical and systematic uncertainties added in quadrature.}
\label{fig:result_z_0}
\end{figure}
When the same procedure is applied to the re{\-}con{\-}structed data events, the obtained event yields---represented by the blue lozenges in Fig.~\ref{fig:ratio_optical_foil_MC}---exhibit a dependence on the target $z$ position. This behavior is
expected due to the variation of the cross section as a~function of the electron scattering angle, because foils at different positions have different acceptances, depending on the mean value of the scattering angle. The observed $z$ dependence of the event yields is in a good agreement with a linear function $(\chi^2/\text{NDF}=0.35)$ and a correction is applied to the data. Note that this behavior is absent in the MC event yields (the red dots in Fig.~\ref{fig:ratio_optical_foil_MC}), because the MC simulation takes into account differences in the acceptance for individual foils.
In the data analysis, we relied on the reconstructed target $z$ position to identify events coming from each of the nine carbon foils, applying the selections represented by the shaded regions in Fig.~\ref{fig:optical_z_distribution}. The selected events were then corrected by using the acceptance matrices computed at $z=0$~cm. To determine the sensitivity to this approximation, we repeated the same study using nine different matrices (one for each carbon foil) and found a negligible variation, as shown in Fig~\ref{fig:result_z_0}. The obtained event yields are subject to the systematic uncertainties due to the $z$-position selection applied to identify events coming from individual foils.

The results obtained using the acceptance method for all targets considered---C, Ti, Ar, and Al---turn out to be in fairly good agreement with those obtained within the yield-ratio method. For carbon, the two methods yield the cross sections differing typically by less than 5\%. For aluminum, the differences typically do not exceed 10\%, as will be discussed in Sec.~\ref{sec:Results}.

It is not unexpected that the differences for aluminum are larger than those for carbon. The Al target is thicker than the C one---0.889 g/cm$^2$ vs 0.167 g/cm$^2$, respectively---and is positioned at $z=-12.5$~cm compared with $z = 0$~cm for carbon. As a consequence, the acceptance-method result for aluminum is subject to larger systematic uncertainties due to the acceptance corrections, of the order of 10\%, which are difficult to precisely estimate. They stem from a broader range of the interaction-vertex positions, and the target's location away from the region of the optimal spectrometer acceptance and beyond the $z$ range characterized by the optical target studies, $-10\leq z\leq 10$ cm.

In this article, we compare the yield-ratio and acceptance methods of extracting the cross section for aluminum, as for this target a large body of available experimental data makes the MC simulation particularly reliable. The observed fairly good agreement supports our conclusion drawn based on the outcome of variations in the Monte Carlo model that the cross section serving as input to the MC simulations does not affect significantly the extracted double-differential cross section. In Sec.~\ref{sec:Uncertainties}, we discuss in detail the systematic uncertainties only for the yield-ratio method, the main approach applied in this analysis.

\section{Uncertainty Analysis}\label{sec:Uncertainties}
The total systematic uncertainty in this analysis was estimated by adding in quadrature the contributions listed in Table~\ref{tab:syst}. Each of the uncertainties was considered completely uncorrelated. We determined the cuts ensuring that there are no dependencies on kinematic variables and, therefore, all the uncertainties affects only the normalization of the extracted cross sections. The kinematic cuts used in the analysis were varied by $\pm$10\% or by the resolution of the variable under consideration.

Because the results obtained depend on the Monte Carlo calculation, it is important to estimate uncertainties resulting from its inputs. To determine the uncertainties related to the target position, we performed the simulation with the inputs for the beam's and spectrometer's $x$ and $y$ offsets varied within uncertainties, and we recomputed the optical transport matrix varying the three quadrupole magnetic fields, one at the time. Each of these runs was compared with the reference run, and the corresponding differences were summed in quadrature to give the total systematic uncertainty due to the Monte Carlo. That summed uncertainty value varied from 1.1\% to 2.2\%, based on the momentum setting for each of the run, and was the largest single source of systematic error. The statistical uncertainty varied between 1.5\% and 2.0\% across the considered kinematical regions.

\begin{table}
\caption{\label{tab:syst}Contributions to systematic uncertainties in the yield-ratio method for aluminum and argon.}
\begin{ruledtabular}
\begin{tabular}{l l c c c c}
	&				&					& Al 	      & Ar          \\
\hline				
a. & Beam energy & 					& 0.1\% & 0.1\% \\
b. & Beam charge &					& 0.3\% & 0.3\% \\
c. & Beam $x$ offset &					& $<1.0$\% & $<0.8\%$ \\
d. & Beam $y$ offset &					& $<1.0$\% & $<0.9\%$ \\
e. & HRS $x$ offset &					& $<0.8$\% & $<1.0\%$ \\
f. & HRS $y$ offset &					& $<0.6$\% & $<0.8\%$ \\
g. & Optics (q1, q2, q3) &				& $<1.8$\% & $<1.0\%$ \\
h. & Target thickness/density/lenght & 				& $0.2$\% & $0.7\%$ \\
i. & Acceptance cut $(\theta,\phi,dp/p)$ & 	& $<1.0$\% & $<2.4\%$ \\
\phantom{1. i. }&(i) $dp$ acceptance cut & 			& $<0.32$\% & - \\
\phantom{1. i. }&(ii) $\theta$ acceptance cut &			& $<0.32$\% & - \\
\phantom{1. i. }& (iii) $\phi$ acceptance cut &			& $<0.79$\% & - \\
\phantom{1. i. }& (iv) $z$ acceptance cut & 			& $<0.45$\% & - \\
j. & Calorimeter cut & 			        & $<0.02$\% & $<0.02\%$ \\
k. & \v{C}erenkov cut &				& $<0.12$\% & $<0.07\%$ \\
l. & Cross section model &				& $<0.2$\%& $<1.3\%$ \\
m. & Radiative and Coulomb corr. & 		        & 1.0\%--1.3\% & 1.0\%--1.3\% \\
\hline
& Total systematic uncertainty &			& 1.7\%--2.7\% & 1.8\%--3.0\% \\
\end{tabular}
\end{ruledtabular}
\end{table}

The systematic uncertainty due to the cuts on the calorimeter and \v{C}erenkov detector was calculated in a similar way, by varying the cut by a small amount and calculating the difference with respect to the nominal cut. Given the already high efficiency of these cuts, this resulted in a very small contribution to the uncertainty. The uncertainty due to the acceptance cuts on the angles and on ${dp/p}$ was calculated in the same way. We included a fixed uncertainty relative to the beam charge and beam energy as in previous work on C and Ti~\cite{Dai:2018xhi}.

The measured cross section is also corrected for the effects from internal and external radiative processes. Internal radiative process are vacuum polarization, vertex corrections, and internal bremsstrahlung. External radiative processes refer to electrons losing energy while passing through material in the target. We applied the radiative corrections using the approach of Whitlow~\cite{Whitlow:1990}, which follows the recipe of Dasu~\cite{Dasu} and is based on the peaking approximation~\cite{Tsai:1969}. This approach is subject to theoretical uncertainties and depends on the cross-section model. We consider a fixed 1\% uncertainty due to the theoretical model for the radiative corrections over the full kinematic range. To account for the cross-section model dependence---the same for both the yield-ratio and acceptance methods---we added an additional uncertainty (fully uncorrelated), estimated by computing the difference in the final double differential cross section when the cross section model is rescaled by $\sqrt{(Q^2)}/2$, with $Q^2$ being the four-momentum transfer squared.
Coulomb corrections were included in the local effective momentum approximation, following Ref.~\cite{Aste:2005wc}. A 10\% uncertainty associated with the Coulomb potential was included as systematic uncertainty.

Near the quasielastic peak, there is a non-negligible contribution of the elastic cross section to the inclusive cross section, through the radiative processes. To estimate the corresponding uncertainty, we increased the tail of the elastic contribution by 20\%, recalculated the radiative correction, and used its difference with respect to the reference correction as an estimate of the corresponding systematic uncertainty. Finally, we included a target thickness uncertainty.

\begin{figure}
\centering
\includegraphics[width=0.9\columnwidth]{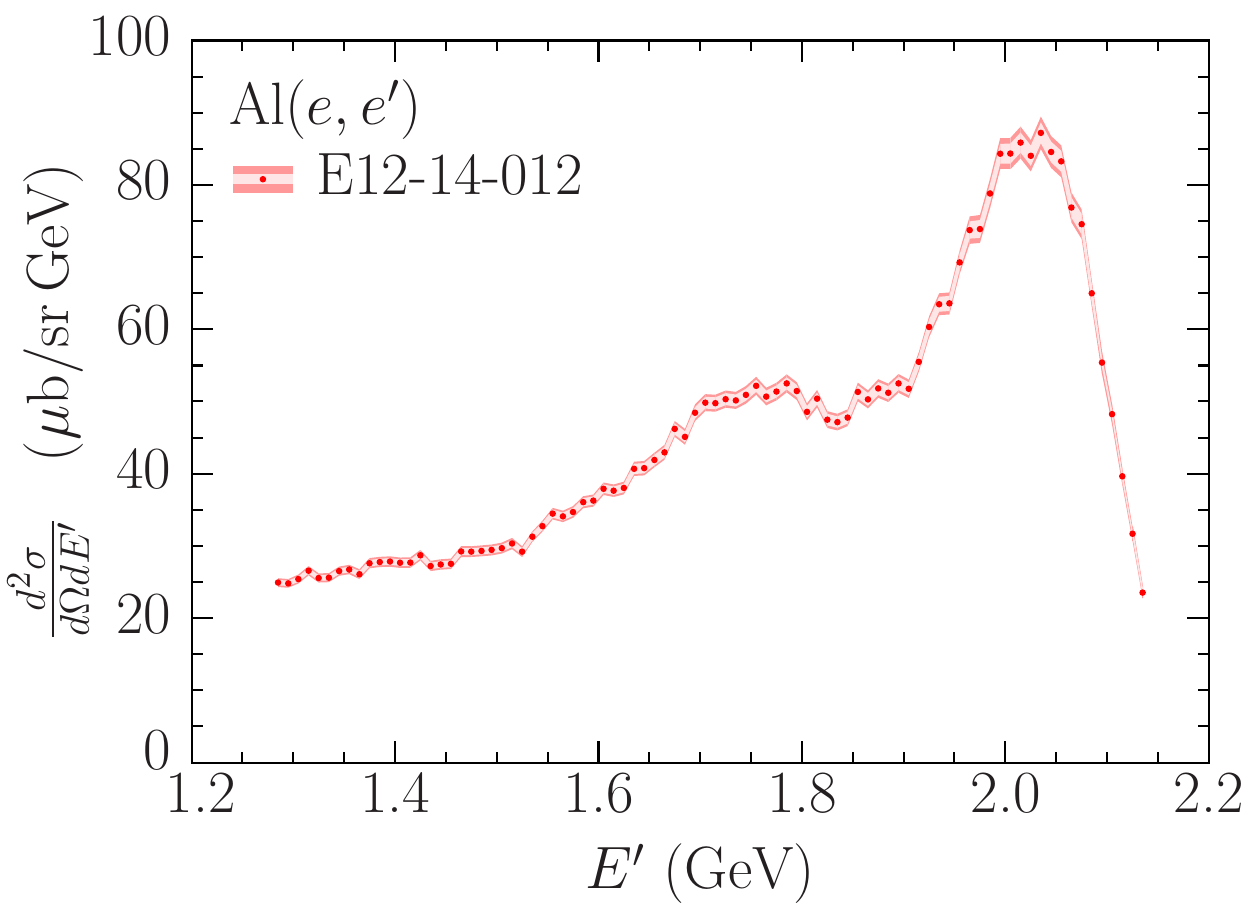}
\caption{(color online). Double-differential cross section extracted for inclusive electron scattering off the Al-7075 target at a beam energy of 2.222~GeV and a scattering angle of $15.54^\circ$. The inner and outer uncertainty bands correspond to statistical and total uncertainties, respectively.}
\label{fig:xsec_Al}
\end{figure}

\section{Experimental Results}\label{sec:Results}

The cross section for inclusive scattering of electrons on the Al-7075 target, extracted by using the yield-ratio method and normalized per nucleus, is shown in Fig.~\ref{fig:xsec_Al}. Its total uncertainties---represented by the outer bands---are obtained by adding in quadrature statistical and systematic uncertainties. As in the case of the previously reported results~\cite{Dai:2018gch,Dai:2018xhi}, the aluminum measurement is very precise and limited by the systematic uncertainties.

\begin{figure}
\centering
\includegraphics[width=0.9\columnwidth]{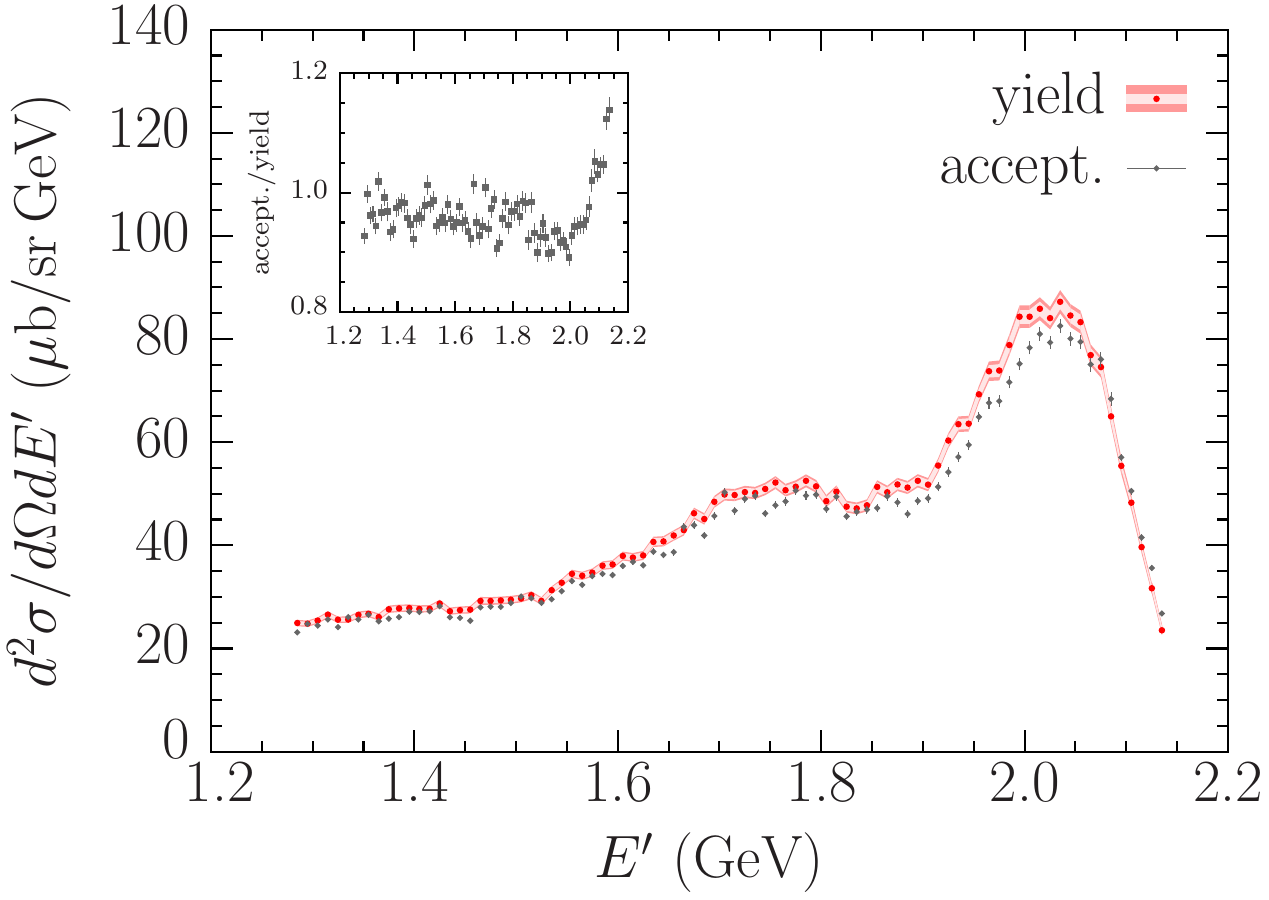}
\caption{(color online). Comparison of the Al$(e,e')$ cross sections extracted using the yield-ratio and acceptance methods, with their ratio presented in the inset. For the yield-method, the inner (outer) bands represent statistical (total) uncertainties. For the acceptance method, only statistical uncertainties are shown.}
\label{fig:compare_Al_yield_acceptance}
\end{figure}

As a cross-check, we also extracted the Al cross section by using the acceptance method. Figure~\ref{fig:compare_Al_yield_acceptance} shows that the results obtained using the two methods are in fairly good agreement. Note that in the acceptance method, the error bars represent the statistical uncertainties only. As discussed in Sec.~\ref{sec:Uncertainties}, systematic uncertainties of the acceptance result are estimated to be of the order of 10\%, three to four times larger than the total uncertainties of the yield-ratio method, and their precise determination would be difficult. The acceptance result is useful nevertheless, for the following reasons.
The agreement between the yield-ratio and acceptance results---observed for the carbon, aluminum, titanium, and argon targets---provides a consistency check between the yield and acceptance methods. It also further corroborates that the yield-ratio results exhibit only weak dependence on the input cross section used in the Monte Carlo simulation to correct the data for efficiency and acceptance, in consistence with direct estimates performed for carbon and aluminum. This issue is particularly important in the case of the titanium and argon targets, where the cross-section simulations cannot be validated against existing data. Note that the radiative corrections applied in both methods are the same, and do depend on the input cross section.

\begin{figure}
\centering
\includegraphics[width=0.9\columnwidth]{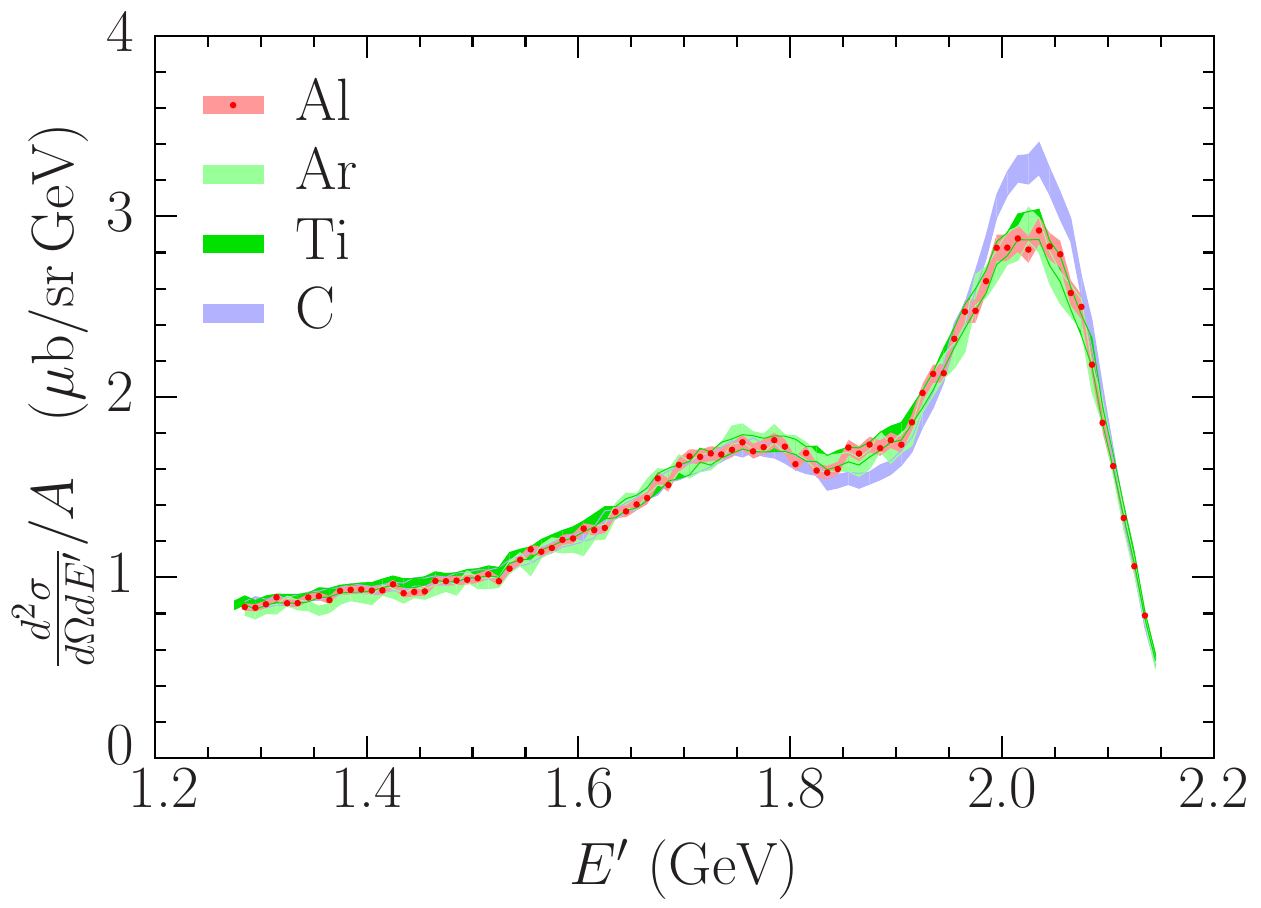}
\caption{(color online). Comparison of the cross sections per nucleon for aluminum, argon~\cite{Dai:2018gch}, titanium~\cite{Dai:2018xhi}, and carbon~\cite{Dai:2018xhi} measured at beam energy 2.222~GeV and scattering angle $15.54^\circ$. The average nucleon number for every target is calculated according to the natural abundances of isotopes, see details in the text. The bands represent the total uncertainties.}
\label{fig:compare_Al_A}
\end{figure}

To illustrate how nuclear effects affect different interaction channels, in Fig.~\ref{fig:compare_Al_A} we compare the per-nucleon cross sections for aluminum, argon, titanium, and carbon. It is important to note that all the results are extracted using the yield-ratio method, following the same procedure, described in Sec.~\ref{sec:yield}. While for every target we account for the abundances of naturally occurring isotopes~\cite{Tuli:2011}, this effect is relevant only for the Al target. It is a consequence of the non-negligible contributions of elements heavier than $^{27}_{13}$Al to the Al-7075 alloy, detailed in Table~\ref{tab:alcomp}. At the considered kinematics, corresponding to the beam energy 2.222~GeV and scattering angle $15.54^\circ$, the cross sections per nucleon for targets ranging from carbon ($A=12.01$) to titanium ($A=47.92$) turn out to be in very good agreement in the region where different pion production mechanisms dominate. While this finding is by no means obvious---due to asymmetry of the proton and neutron numbers for aluminum, argon, and titanium---it is consistent with the results of Refs.~\cite{O'Connell:1987ag,Sealock:1989nx} at energies $\sim$0.54--1.50~GeV and scattering angles $\sim$37$^\circ$.

The influence of nuclear effects on QE interactions can be better illustrated in terms of the cross sections normalized to the elementary contributions of neutrons and protons that compose the nucleus; that is, the quantity
\begin{equation}
\frac{d^2\sigma}{d\Omega d E^\prime} / [Z\tilde\sigma_{ep} + N\tilde\sigma_{en}],
\label{ratio}
\end{equation}
where $Z$ and $N$ are the proton and neutron numbers, respectively, while $\tilde\sigma_{ep}$ and $\tilde\sigma_{en}$ denote the elastic electron-proton and electron-neutron cross sections stripped of the energy-conserving $\delta$ function~\cite{deForest:1983}. In the following, we use the average neutron numbers calculated according to the natural abundances of isotopes, that is 6.01 for carbon, 21.98 for argon, and 25.92 for titanium~\cite{Tuli:2011}. For the aluminum target, we employ $Z = 14.26$ and $N = 15.58$, due to its composition listed in Table~\ref{tab:alcomp}.

\begin{figure}
\centering
\includegraphics[width=0.9\columnwidth]{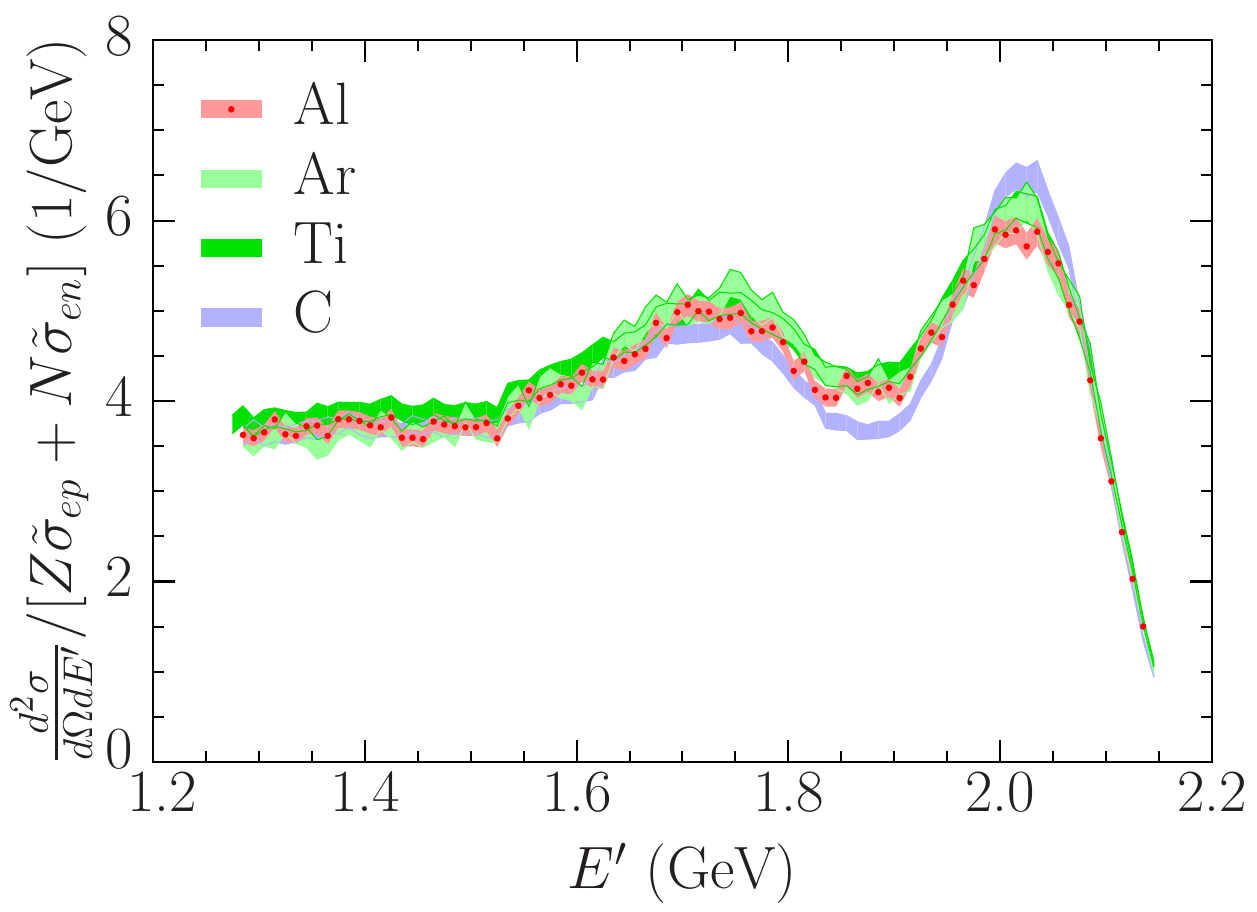}
\caption{(color online). Same as in Fig.~\ref{fig:compare_Al_A} but for the cross sections normalized by the combination of the elementary cross sections according to Eq.~\eqref{ratio}.}
\label{fig:compare_Al_cs}
\end{figure}

As shown in Fig.~\ref{fig:compare_Al_cs}, the results for titanium and argon are, within uncertainties, identical in the QE peak, but they differ from both those for carbon and aluminum. Near the maximum of the QE peak, the cross section defined in Eq.~\eqref{ratio} is lower by $\sim$4\% for aluminum, and higher by $\sim$5\% for carbon, than those for argon and titanium. In the dip region, the results for aluminum (carbon) are lower by $\sim$2\% ($\sim$13\%) compared with those for argon and titanium.

In QE scattering, the cross sections normalized according to Eq.~\eqref{ratio} exhibit very weak target dependence only in the region of high $E'$, corresponding to low energy transfers, as shown in Fig.~\ref{fig:compare_Al_cs}. This is, however, not the case in the QE peak's maximum and for lower $E'$, where the energy transferred by electrons to the nucleus is sufficiently high to probe deeply bound states and also to induce two-nucleon knockout.

The observed differences in the dependence on the atom{\-}ic number of various interaction mech{\-}a{\-}nisms---pre{\-}vi{\-}ous{\-}ly reported in Refs.~\cite{OConnell:1984qim,Sealock:1989nx,Baran:1988tw}---can be expected to provide important clues for building models of nuclear effects valid over broad kinematic regimes and able to describe a~range of targets. Such models are of great importance to long-baseline neutrino-oscillation experiments.

%
%
%
\section[Scaling and A-dependence]{Scaling and \texorpdfstring{$A$}{A}-dependence}\label{sec:scaling}
The scaling analysis allows us to compare inclusive elec{\-}tron-scattering data taken in different kinematic conditions and using different targets.

Scaling of first kind, or $y$-scaling, is observed in the kinematic region of large momentum transfer, $\norm{q}$, and energy transfer
$\omega < \sqrt{\norm{q}^2 + m^2} - m$, in which the beam particle interacts with individual nucleons and the dominant
reaction mechanism is quasielastic scattering~\cite{Sick:1980,Day:1990mf}. Under these conditions, the target response, which in general depends on both
momentum and energy transfers, reduces to a function of the single variable $y = y(\norm{q},\omega)$, defined by the equation
\begin{equation}
\begin{split}
\label{def:y}
\omega + M_A  & = \sqrt{ y^2 + (M_A - m + E_{\rm min})^2  }\\
&\quad + \sqrt{ (y + \norm{q})^2 + m^2 }.
\end{split}
\end{equation}
Here, $m$ and $M_A$ are the nucleon mass and the target-nucleus mass, respectively, while $E_{\rm min}$ denotes the nu{\-}cle{\-}on-knockout threshold. The scaling variable $y$, having the dimension of energy, is simply related to the longitudinal component of the initial momentum of the struck nucleon,
$k_\parallel = {\bf k} \cdot {\bf q}/|{\bf q}|$.  The scaling function $F(y)$ is determined from the measured cross section, $ \sigma^{\rm exp}$ through
\begin{equation}
F(y) = K\frac{\sigma^{\rm exp}}{Z \tilde{\sigma}_{ep} + N  \tilde{\sigma}_{en}},
\end{equation}
with $K$ being a kinematic factor.

\begin{figure}
\centering
\includegraphics[width=1.0\columnwidth]{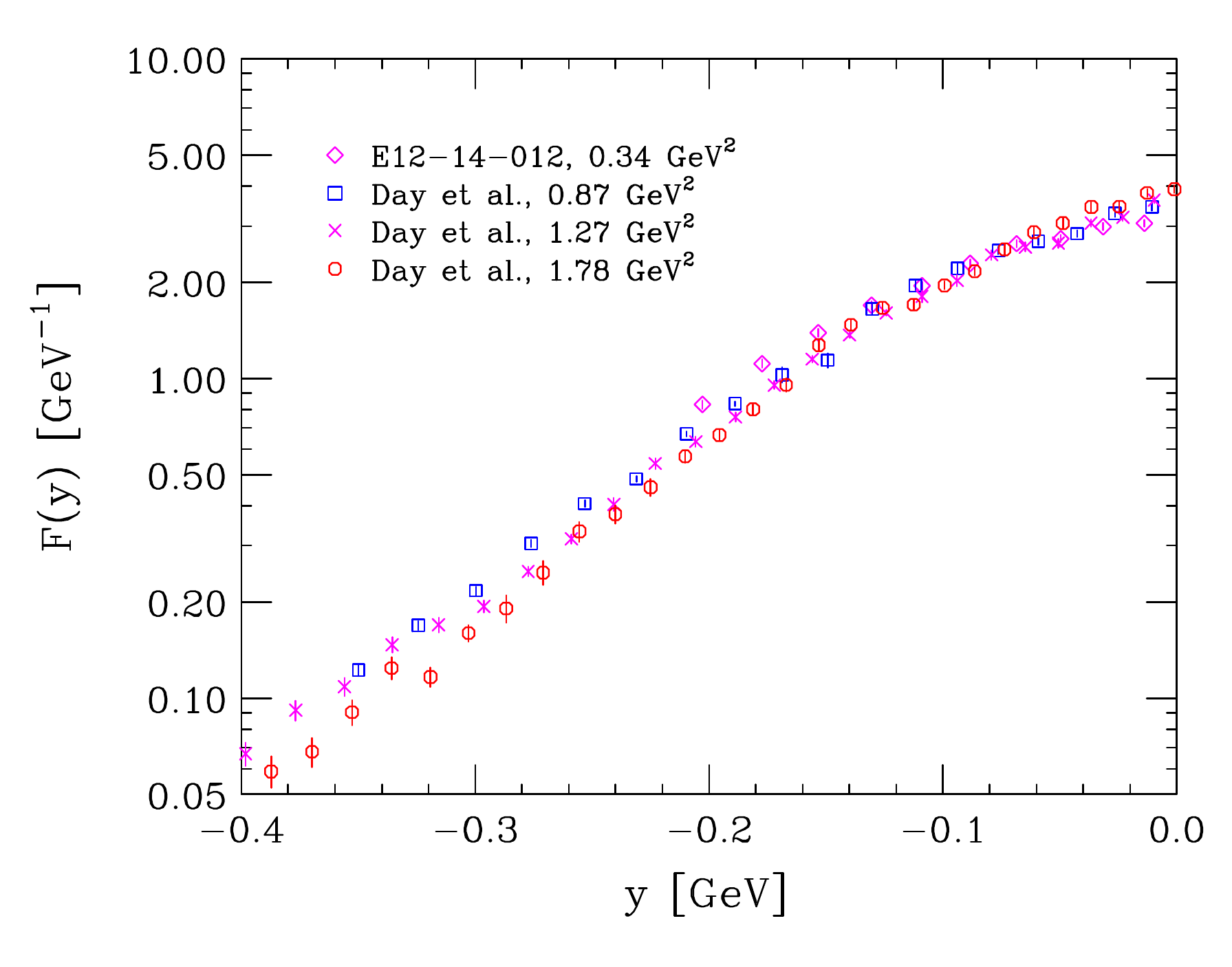}
\caption{(color online) Comparison between the scaling function of aluminum obtained from the E12-14-012 data (this work), represented by diamonds, and those obtained from the data of Day~{\it et~al.}~\cite{Day}.
The data are labeled according to the value of $Q^2$ corresponding to quasielastic kinematics.}
\label{fig:yscaling_1}
\end{figure}

\begin{figure}
\centering
\includegraphics[width=1.00\columnwidth]{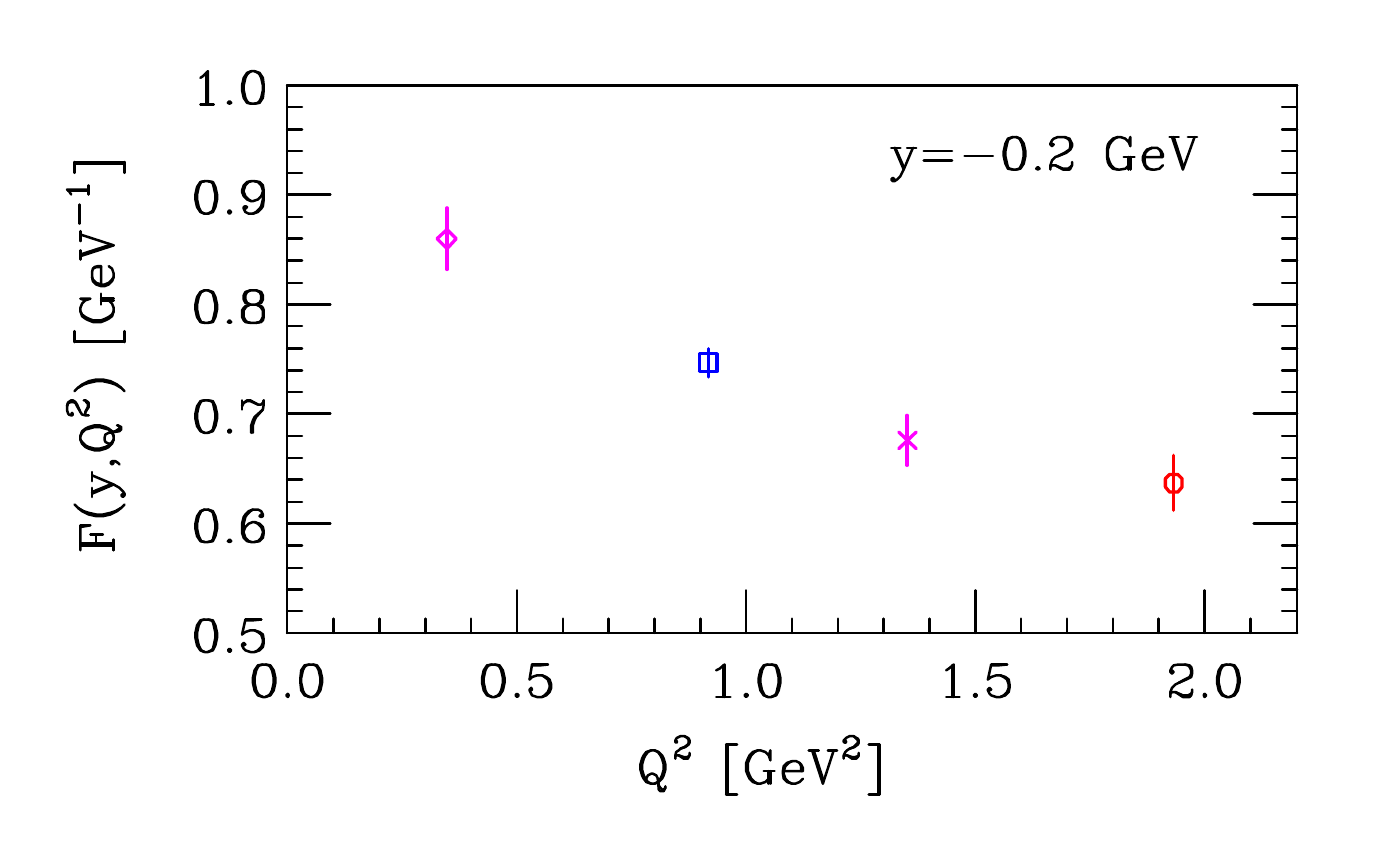}
\caption{(color online) $Q^2$-dependence of the scaling function $F(y,Q^2)$ obtained from the cross section displayed in Fig.~\ref{fig:xsec_Al} and from the data reported in Ref.~\cite{Day} at fixed $y = -0.2$ GeV. The meaning of the symbols is the same as in Fig.~\ref{fig:yscaling_1}. }
\label{fig:yscaling_2}
\end{figure}

The results of the $y$-scaling analysis of aluminum data are illustrated in Figs.~\ref{fig:yscaling_1} and~\ref{fig:yscaling_2}. The scaling function computed using the cross section displayed in Fig.~\ref{fig:xsec_Al} and the average proton and neutron numbers from Table~\ref{tab:alcomp} is compared with those obtained from the  $^{27}_{13}$Al data of Day {\it et al.}~\cite{Day}.  The cross sections of Ref.~\cite{Day} were measured in SLAC at a~fixed beam energy $E=3.595$~GeV  and scattering angles $16^\circ$, $20^\circ$, and $25^\circ$, with the values of $Q^2$ corresponding to quasielastic kinematics being 0.87, 1.27 and 1.78 GeV$^2$, respectively.
Figure~\ref{fig:yscaling_1} shows that the agreement is very good at $y \approx 0$, corresponding to quasielastic kinematics, $\omega \approx Q^2/2m$, while significant scaling violations occur at large negative $y$. Figure~\ref{fig:yscaling_2}, showing that the scaling limit is approached from above, confirms that these scaling violations arise mainly from the effects of FSI between the knocked-out nucleon and the spectator particles. It is apparent that the data point corresponding to our experiment fits very well into the pattern described by the SLAC data~\cite{Day}.

While the occurrence of $y$-scaling simply reflects the dom{\-}i{\-}nance of quasielastic single-nucleon knockout, a~more general form of scaling, dubbed scaling of the second kind, permits a global analysis, combining data corresponding to different targets~\cite{Donnelly:1999}. The definitions of the dimensionless scaling variable, $\psi$, and scaling function, $f(\psi)$, involve a momentum scale---somewhat misleadingly called Fermi momentum---providing a parametrization of the target-mass dependence of the measured cross sections.

The nuclear Fermi momentum is a well-defined quantity only within the Fermi gas model, and can only be obtained by applying this model to the description of electron-scattering data. On the other hand, the determination of the Fermi momentum from optimization of the scaling analysis around $\psi = 0$ involves two important issues. (i) In principle, the description of nuclei with significant neutron excess requires two different Fermi momenta: for protons and for neutrons. (ii) Significant scaling violations are known to occur, and should be properly taken into account to achieve an accurate determination of the Fermi momentum.

\begin{figure}
\centering
\includegraphics[width=0.95\columnwidth]{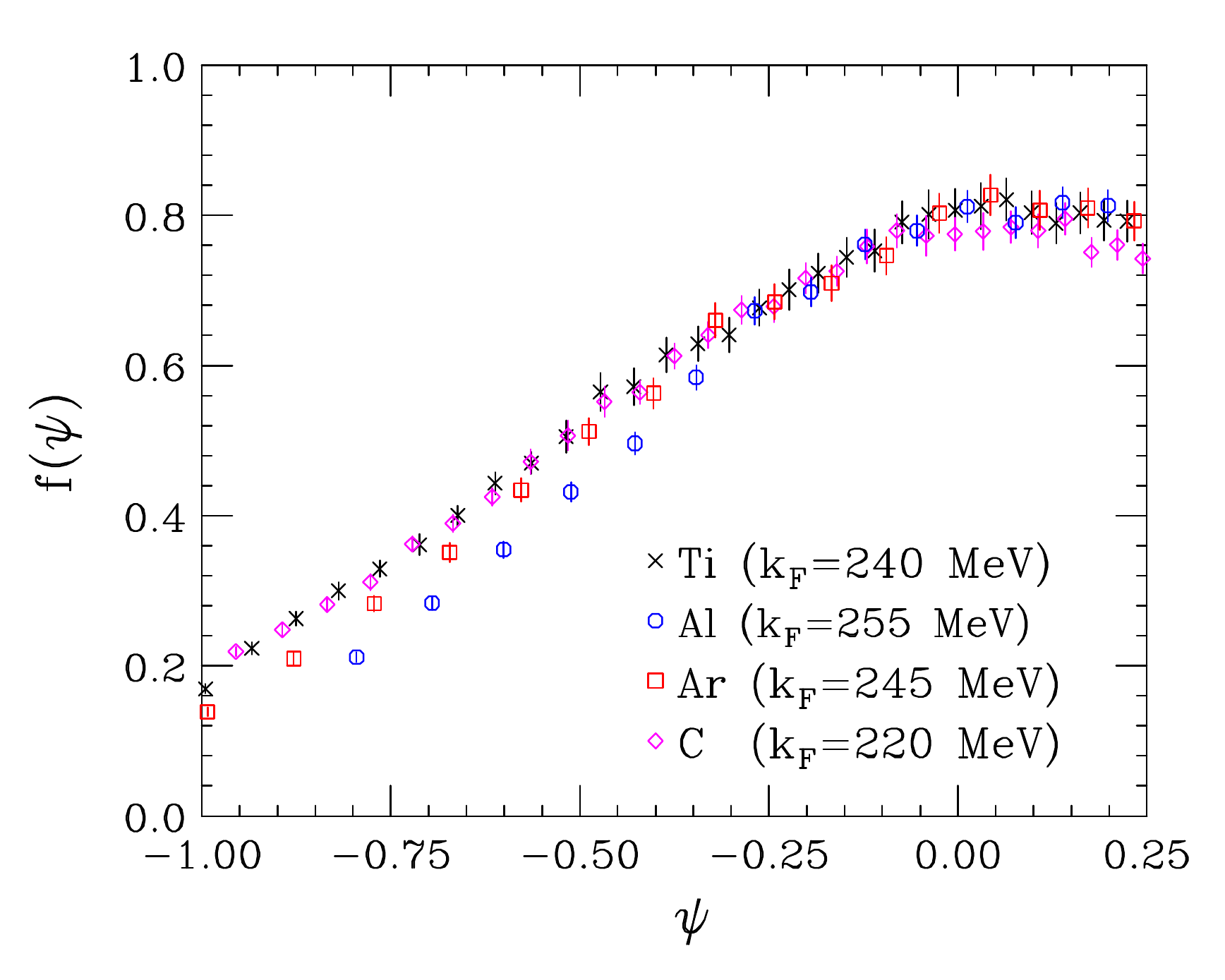}
\caption{(color online) Scaling functions of second kind, obtained from the inclusive cross sections measured by the E12-14-012 experiment using carbon, aluminum, argon and titanium targets.}
\label{fig:superscaling}
\end{figure}

To study scaling of second kind, we adopt for carbon the Fermi momentum $k_F=220$ MeV, obtained from the independent analysis of Moniz {\it et al.}~\cite{Moniz:1971mt}, performed consistently within the Fermi gas model. As the measurements reported in Ref.~\cite{Moniz:1971mt} did not extend to the heavier targets discussed here, in those cases  we estimate the $k_F$ values from the scaling behavior around $\psi=0$. Figure~\ref{fig:superscaling} illustrates that the inclusive cross sections measured by the E12-14-012 experiment exhibit scaling of the second kind, when $k_F$ values of 255, 245, and 240 MeV  are taken for Al, Ar, and Ti, respectively. This finding provides an additional cross-check of the self-consistency of the measurements performed in the JLab experiment E12-14-012.

\section{Summary and Conclusions}\label{sec:Summary}

We have reported on the measurements of the cross sections for inclusive electron scattering over a broad range of energy transfers, extending from the particle-emission threshold to above the excitation of the first hadronic resonance. These high-precision data were taken at Jefferson Lab in Hall~A for a beam energy of $E=2.222$ GeV and electron-scattering angle $\theta = 15.54^\circ$ from four nuclear targets: carbon, aluminum, argon, and titanium. The reported results give a unique opportunity to validate nuclear models employed in Monte Carlo simulations of precise long-baseline neutrino-oscillation experiments, and to assess their contribution to uncertainties of the oscillation analysis in a rigorous manner.

We find (see Fig.~\ref{fig:compare_Al_A}) that the per-nucleon responses for the four targets considered are strikingly similar over the entire energy-transfer range ($ 0.05 < \omega < 0.90$ GeV), save at the maximum of the quasielastic peak and the dip region. At the kinematics from the maximum of the quasielastic peak to the onset of the $\Delta$ resonance, the result for carbon stands apart from those for aluminum, argon, and titanium. This finding shows that the momentum and energy distribution of nucleons in the nuclear ground state and final-state interactions---inducing the `Doppler' broadening of the scattered electron's final energy---in carbon is not as pronounced as for the heavier nuclei.
When accounting is made for the number of protons and neutrons in each nucleus, this feature does not disappear, as can be seen in Fig.~\ref{fig:compare_Al_cs}.

When the extracted aluminum data are presented in terms of the $y$-scaling analysis (Fig.~\ref{fig:yscaling_1}) along with the higher-$Q^2$ data from SLAC, the set behaves as expected, and the scaling behavior is clearly observed at the kinematics corresponding to the quasifree peak. While in the absence of FSI, the scaling  function $F(y)$ is expected to converge from below with increasing $Q^2$, the effect of FSI---falling with $Q^2$---leads it to converge from above. These new data fit this pattern (Fig.~\ref{fig:yscaling_2}).

Taken together, this data set will allow us to predict by interpolation the electromagnetic nuclear responses for nuclei between $A = 12$ and 48. Of particular interest will be oxygen, because water serves as the target and radiator in the large \v{C}erenkov detector of T2K~\cite{Abe:2018wpn}, and chlorine, because polyvinyl chloride composes the detectors of NOvA~\cite{Talaga:2016rlq}.

\begin{acknowledgments}
We acknowledge the outstanding support from the Jefferson Lab Hall A technical staff, target group and Accelerator Division. This experiment was made possible by Virginia Tech and the National Science Foundation under CAREER grant No. PHY$-$1352106. This work was also supported by the DOE Office of Science, Office of Nuclear Physics, contract DE-AC05-06OR23177, under which Jefferson Science Associates, LLC operates JLab, DOE contracts DE-FG02-96ER40950, DE-AC02-76SF00515, and DE{-}SC0013615.
\end{acknowledgments}
%
%
%
%
%

\end{document}